\documentstyle[12pt,epsf]{article}

\newlength{\pubnumber} 

\catcode`\@=11
\@addtoreset{equation}{section}

\def\section{\@startsection{section}{1}{\z@}{3.5ex plus 1ex minus .2ex}
 {2.3ex plus .2ex}{\large\bf}}
\def\subsection{\@startsection{subsection}{2}{\z@}{2.3ex plus .2ex}
 {2.3ex plus .2ex}{\bf}}

\begin{document}

\begin{titlepage}
\samepage{
\setcounter{page}{1}
\rightline{ACT-13/00}
\rightline{CERN-TH/2000-266}
\rightline{CTP-TAMU-30/00}
\rightline{\tt hep-ph/0009338}
\rightline{September 2000}
\vfill
\begin{center}
 {\Large \bf Flat Directions in Flipped $SU(5)$ I: All-Order Analysis\\ }
\vfill
\vskip .4truecm
\vfill {\large
        G.B. Cleaver$^{1,2}$\footnote{gcleaver@rainbow.physics.tamu.edu},
        J. Ellis$^{3}$\footnote{john.ellis@cern.ch} and
        D.V. Nanopoulos$^{1,2,4}$\footnote{dimitri@soda.physics.tamu.edu}}
\\
\vspace{.12in}
{\it $^{1}$ Center for Theoretical Physics,
            Dept.\  of Physics, Texas A\&M University,\\
            College Station, TX 77843, USA\\}
\vspace{.06in}
{\it $^{2}$ Astroparticle Physics Group,
            Houston Advanced Research Center (HARC),\\
            The Mitchell Campus,
            Woodlands, TX 77381, USA\\}
\vspace{.06in}
{\it$^{3}$  Theory Division, CERN, 
            CH 1211 Geneva 23, Switzerland\\}
\vspace{.06in}
{\it$^{4}$  Academy of Athens, Chair of Theoretical Physics, 
            Division of Natural Sciences,\\
            28 Panepistimiou Avenue, Athens 10679, Greece\\}
\vspace{.025in}
\end{center}
\vfill
\begin{abstract}
We present a systematic classification of field directions 
for the string--derived flipped $SU(5)$ model that are $D$-- and
$F$--flat to all orders.
Properties of the flipped $SU(5)$ model with field values in these
directions
are compared to those associated with other flat directions that have
been shown to be $F$--flat to specific finite orders in the
superpotential. 
We discuss the phenomenological Higgs spectrum, and quark and charged--lepton
mass textures.
\end{abstract}
\smallskip}
\end{titlepage}

\setcounter{footnote}{0}

% ========================= DEFINITIONS ===================================
% latex commands

% abbreviations

\def\beq{\begin{equation}}
\def\eeq{\end{equation}}
\def\beqn{\begin{eqnarray}}
\def\eeqn{\end{eqnarray}}
\def\no{\noindent }
\def\nolabel{\nonumber }

\def\fsu{flipped $SU(5)$ }
\def\NA{non--Abelian }

\def\gsim{{\buildrel >\over \sim}}
\def\lsim{{\buildrel <\over \sim}}

\def\ie{i.e., }
\def\eg{{\it e.g.}}
\def\eq#1{eq.\ (\ref{#1})}
\def\eqs#1{eqs.\ (\ref{#1})}

\def\lt{<}

\def\slash#1{#1\hskip-6pt/\hskip6pt}
\def\slk{\slash{k}}

\def\dag{\dagger}
\def\qandq{\quad {\rm and} \quad} 
\def\qand{\quad {\rm and} } 
\def\andq{ {\rm and} \quad } 
\def\qwithq{\quad {\rm with} \quad} 
\def\qwith{ \quad {\rm with} } 
\def\withq{ {\rm with} \quad} 

\def\fhalf{\frac{1}{2}}
\def\fsqrt{\frac{1}{\sqrt{2}}}
\def\half{{\textstyle{1\over 2}}}
\def\third{{\textstyle {1\over3}}}
\def\quarter{{\textstyle {1\over4}}}
\def\sixth{{\textstyle {1\over6}}}
\def\m{{\tt -}}
\def\ps{{\tt +}}
\def\pps{\phantom{+}}

\def\zz{$Z_2\times Z_2$ }

\def\Tr{{\rm Tr}\, }
\def\tr{{\rm tr}\, }

\def\GeV{\,{\rm GeV}}
\def\TeV{\,{\rm TeV}}

\def\lam#1{\lambda_{#1}}
\def\non{\nonumber}
\def\fd{flat direction }
\def\smgg{ $SU(3)_C\times SU(2)_L\times U(1)_Y$ }
\def\SM{Standard--Model }
\def\SUSY{supersymmetry }
\def\SSSM{supersymmetric standard model}
\def\MSSM{minimal supersymmetric standard model}
\def\obs{{\rm observable}}
\def\sig{{\rm singlets}}
\def\hid{{\rm hidden}}
\def\MS{M_{string}}
\def\Ms{$M_{string}$}
\def\gs{g_{\rm string}}
\def\gst{g^2_{\rm string}}
\def\MP{M_{P}}
\def\MPt{M^2_{P}}

\def\vev#1{\langle #1 \rangle}
\def\mvev#1{|\langle #1 \rangle |^2}

\def\UA{U(1)_{\rm A}}
\def\QA{Q^{(\rm A)}}
\def\mssm{SU(3)_C\times SU(2)_L\times U(1)_Y} 

\def\KM{Ka\v c--Moody }

\def\y{\,{\rm y}}
\def\l{\langle}
\def\r{\rangle}
\def\o#1{\frac{1}{#1}}

\def\zi{z_{\infty}}

\def\hb#1{\bar{h}_{#1}}
\def\bh#1{\bar{h}_{#1}}
\def\Htw{{\tilde H}}
\def\chibar{{\overline{\chi}}}
\def\qbar{{\overline{q}}}
\def\ibar{{\overline{\imath}}}
\def\jbar{{\overline{\jmath}}}
\def\Hbar{{\overline{H}}}
\def\Qbar{{\overline{Q}}}
\def\abar{{\overline{a}}}
\def\alphabar{{\overline{\alpha}}}
\def\betabar{{\overline{\beta}}}
\def\tautwo{{ \tau_2 }}
\def\thetatwo{{ \vartheta_2 }}
\def\thetathree{{ \vartheta_3 }}
\def\thetafour{{ \vartheta_4 }}
\def\ttwo{{\vartheta_2}}
\def\tthree{{\vartheta_3}}
\def\tfour{{\vartheta_4}}
\def\ti{{\vartheta_i}}
\def\tj{{\vartheta_j}}
\def\tk{{\vartheta_k}}
\def\calF{{\cal F}}
\def\smallmatrix#1#2#3#4{{ {{#1}~{#2}\choose{#3}~{#4}} }}
\def\ab{{\alpha\beta}}
\def\Minv{{ (M^{-1}_\ab)_{ij} }}
\def\ii{{(i)}}
\def\V{{\bf V}}
\def\N{{\bf N}}

% for basis vectors:
\def\bmit{\mathbf}
\def\ob#1{${\bf b}_{#1}$}
\def\b#1{{\bf b}_{#1}}
\def\dv#1{{\bf d}_{#1}}
\def\dvs{{\bf d}}
\def\dvp#1{{\bf d}^{'}_{#1}}
\def\dvpn{{\bf d}^{'}}
\def\dvpp#1{{\bf d}^{(')}_{#1}}
\def\S{{\bf S}}
\def\X{{\bf X}}
\def\I{{\bf I}}
\def\bone{{\mathbf 1}}
\def\obo{${\mathbf 0}$}
\def\bo{{\mathbf 0}}
\def\bs{{\mathbf S}}
\def\bb{{\mathbf b}}
\def\mb{{\mathbf b}}
\def\mS{{\mathbf S}}
\def\bS{{\mathbf S}}
\def\bs{{\mathbf s}}
\def\mX{{\mathbf X}}
\def\bX{{\mathbf X}}
\def\mI{{\mathbf I}}
\def\bI{{\mathbf I}}
\def\balpha{{\mathbf \alpha}}
\def\bbeta{{\mathbf \beta}}
\def\bgamma{{\mathbf \gamma}}
\def\bxi{{\mathbf \xi}}
\def\malpha{{\mathbf \alpha}}
\def\mbeta{{\mathbf \beta}}
\def\mgamma{{\mathbf \gamma}}
\def\mxi{{\mathbf \xi}}
\def\bphi{\overline{\Phi}}
\def\ba{\bar{\alpha}}

\def\eps{\epsilon}

\def\t#1#2{{ \Theta\left\lbrack \matrix{ {#1}\cr {#2}\cr }\right\rbrack }}
\def\C#1#2{{ C\left\lbrack \matrix{ {#1}\cr {#2}\cr }\right\rbrack }}
\def\tp#1#2{{ \Theta'\left\lbrack \matrix{ {#1}\cr {#2}\cr }\right\rbrack }}
\def\tpp#1#2{{ \Theta''\left\lbrack \matrix{ {#1}\cr {#2}\cr }\right\rbrack }}
\def\l{\langle}
\def\r{\rangle}

% Phi symbols outside of $ $
\def\op#1{$\phi_{#1}$}
\def\opp#1{$\phi^{'}_{#1}$}
\def\opb#1{$\overline{\phi}_{#1}$}
\def\opbp#1{$\overline{\phi}^{'}_{#1}$}
\def\oppb#1{$\overline{\phi}^{'}_{#1}$}
\def\oppx#1{$\phi^{(')}_{#1}$}
\def\opbpx#1{$\overline{\phi}^{(')}_{#1}$}

\def\opi{\op{i}}
\def\opil{\op{i\in\{1,2,3,4\}}}

\def\oP#1{$\Phi_{#1}$}
\def\oPp#1{$\Phi^{'}_{#1}$}
\def\oPb#1{$\overline{\Phi}_{#1}$}
\def\oPbp#1{$\overline{\Phi}^{'}_{#1}$}
\def\oPpb#1{$\overline{\Phi}^{'}_{#1}$}
\def\oPpx#1{$\Phi^{(')}_{#1}$}
\def\oPbpx#1{$\overline{\Phi}^{(')}_{#1}$}

\def\oh#1{$h_{#1}$}
\def\ohb#1{${\bar{h}}_{#1}$}
\def\ohp#1{$h^{'}_{#1}$}

\def\oQ#1{$Q_{#1}$}
\def\odc#1{$d^{c}_{#1}$}
\def\ouc#1{$u^{c}_{#1}$}

\def\oL#1{$L_{#1}$}
\def\oec#1{$e^{c}_{#1}$}
\def\oNc#1{$N^{c}_{#1}$}

\def\oH#1{$H_{#1}$}
\def\oV#1{$V_{#1}$}
\def\oHs#1{$H^{s}_{#1}$}
\def\oVs#1{$V^{s}_{#1}$}

\def\oF#1{$F_{#1}$}
\def\oFb#1{$\bar{F}_{#1}$}
\def\of#1{$f_{#1}$}
\def\ofb#1{$\bar{f}_{#1}$}
\def\olc#1{$l^c_{#1}$}
\def\olcb#1{$\bar{l}^c_{#1}$}

\def\mbf{\mathbf}

% Phi symbols inside of $ $
\def\P#1{\Phi_{#1}}
\def\Pp#1{\Phi^{'}_{#1}}
\def\Pb#1{{\overline{\Phi}}_{#1}}
\def\bP#1{{\overline{\Phi}}_{#1}}
\def\Pbp#1{{\overline{\Phi}}^{'}_{#1}}
\def\Ppb#1{{\overline{\Phi}}^{'}_{#1}}
\def\Ppx#1{\Phi^{(')}_{#1}}
\def\Pbpx#1{\overline{\Phi}^{(')}_{#1}}

\def\p#1{\phi_{#1}}
\def\pp#1{\phi^{'}_{#1}}
\def\pb#1{{\overline{\phi}}_{#1}}
\def\bp#1{{\overline{\phi}}_{#1}}
\def\pbp#1{{\overline{\phi}}^{'}_{#1}}
\def\ppb#1{{\overline{\phi}}^{'}_{#1}}
\def\ppx#1{\phi^{(')}_{#1}}
\def\pbpx#1{\overline{\phi}^{(')}_{#1}}

\def\h#1{h_{#1}}
\def\hb#1{{\bar{h}}_{#1}}
\def\hp#1{h^{'}_{#1}}

\def\Q#1{Q_{#1}}
\def\dc#1{d^{c}_{#1}}
\def\uc#1{u^{c}_{#1}}

\def\L#1{L_{#1}}
\def\ec#1{e^{c}_{#1}}
\def\lc#1{l^{c}_{#1}}
\def\Nc#1{N^{c}_{#1}}

\def\H#1{H_{#1}}
\def\V#1{V_{#1}}
\def\Hs#1{{H^{s}_{#1}}}
\def\Vs#1{{V^{s}_{#1}}}

\def\F#1{F_{#1}}
\def\Fb#1{{{\bar{F}}_{#1}}}
\def\bF#1{{{\bar{F}}_{#1}}}

\def\f#1{f_{#1}}
\def\fb#1{{{\bar{f}}_{#1}}}

\def\od#1{$\Delta_{#1}$}
\def\d#1{\Delta_{#1}}
\def\T#1{T_{#1}}
\def\oT#1{$T_{#1}$}

\def\a#1{a_{#1}}
\def\oa#1{$a_{#1}$}
\def\an{a}

\def\ba#1{{\bar a}_{#1}}
\def\oba#1{$\bar{a}_{#1}$}

\def\bap#1{{\bar a}^{'}_{#1}}
\def\obap#1{$\bar{a}^{'}_{#1}$}

\def\ca#1{{\bar a}_{#1}}
\def\oca#1{$\bar{a}_{#1}$}

\def\al{$\vev{\alpha}$}

%================== BLACKBOARD BOLD CHARACTERS ==============================

\def\inbar{\,\vrule height1.5ex width.4pt depth0pt}

\def\IC{\relax\hbox{$\inbar\kern-.3em{\rm C}$}}
\def\IQ{\relax\hbox{$\inbar\kern-.3em{\rm Q}$}}
\def\IR{\relax{\rm I\kern-.18em R}}
 \font\cmss=cmss10 \font\cmsss=cmss10 at 7pt
\def\IZ{\relax\ifmmode\mathchoice
 {\hbox{\cmss Z\kern-.4em Z}}{\hbox{\cmss Z\kern-.4em Z}}
 {\lower.9pt\hbox{\cmsss Z\kern-.4em Z}}
 {\lower1.2pt\hbox{\cmsss Z\kern-.4em Z}}\else{\cmss Z\kern-.4em Z}\fi}

%==============================================================================
\hyphenation{su-per-sym-met-ric non-su-per-sym-met-ric}
\hyphenation{space-time-super-sym-met-ric}
\hyphenation{mod-u-lar mod-u-lar--in-var-i-ant}
%==============================================================================

%========================================================================
%          MACROS FOR REFERENCES
%========================================================================
\def\AEF{A.E. Faraggi}
\def\AP#1#2#3{{\it Ann.\ Phys.}\/ {\bf#1} (#2) #3}
\def\EPJC#1#2#3{{\it The Euro.\ Phys.\ Jour.\/} {\bf C#1} (#2) #3}
\def\JHEP#1#2#3{{\it Jour.\ High\ Ener. Phys.}\/ {\bf #1} (#2) #3}
\def\NPB#1#2#3{{\it Nucl.\ Phys.}\/ {\bf B#1} (#2) #3}
\def\NPBPS#1#2#3{{\it Nucl.\ Phys.}\/ {{\bf B} (Proc. Suppl.) {\bf #1}} (#2) 
 #3}
\def\PLB#1#2#3{{\it Phys.\ Lett.}\/ {\bf B#1} (#2) #3}
\def\PRD#1#2#3{{\it Phys.\ Rev.}\/ {\bf D#1} (#2) #3}
\def\PRL#1#2#3{{\it Phys.\ Rev.\ Lett.}\/ {\bf #1} (#2) #3}
\def\PRT#1#2#3{{\it Phys.\ Rep.}\/ {\bf#1} (#2) #3}
\def\PTP#1#2#3{{\it Prog.\ Theo.\ Phys.}\/ {\bf#1} (#2) #3}
\def\MODA#1#2#3{{\it Mod.\ Phys.\ Lett.}\/ {\bf A#1} (#2) #3}
\def\IJMP#1#2#3{{\it Int.\ J.\ Mod.\ Phys.}\/ {\bf A#1} (#2) #3}
\def\nuvc#1#2#3{{\it Nuovo Cimento}\/ {\bf #1A} (#2) #3}
\def\RPP#1#2#3{{\it Rept.\ Prog.\ Phys.}\/ {\bf #1} (#2) #3}
\def\etal{{\it et al.\/}}
%========================================================================

%============================== SECTION 1 ============================

\setcounter{footnote}{0}

\section{Introduction}
%String--Derived Flipped $SU(5)$}

Over its approximately thirteen--year history~\cite{revit,morefsu5,fsu5a},
the string--derived supersymmetric flipped $SU(5)$  
has become one of the more developed perturbative heterotic string
models~\cite{fsu90,fsu97,fsu98,fsu99a,fsu99b}, and has achieved
several phenomenological successes.
Much of the strength and uniqueness of 
(supersymmetric) flipped $SU(5)$ lies in the fact that, 
unlike conventional GUT models based on $E_6$, $SO(10)$, or 
$SU(5)$ gauge groups, it can be broken to the Standard Model
$SU(3)_C\times SU(2)_L \times U(1)_{Y}$ gauge group without the need of
adjoint or larger Higgs representations. This is important because
it was proven long ago that the presence of massless
adjoint or larger scalar multiplets was inconsistent with $N=1$ or $0$
spacetime supersymmetry
in string models with an underlying level--1 Ka\v c--Moody
algebra~\cite{Reiss}.
In level--1 $SU(5)$, or flipped $SU(5)$, the only allowed massless
representations are   
$\mathbf 1$, $\mathbf 5$, $\bar {\mathbf 5}$, ${\mathbf 10}$, 
and $\bar {\mathbf 10}$. These are not sufficient to break $SU(5)
\rightarrow SU(3)_C \times SU(2)_L \times U(1)_{Y}$, but
are sufficient to break flipped $SU(5) \rightarrow SU(3)_C\times SU(2)_L
\times U(1)_{Y} \rightarrow SU(3)_C \times U(1)_{EM}$~\cite{revit}.

In either conventional or flipped $SU(5)$~\cite{oldfsu5}, a single
generation of 16
matter fields (including
a singlet `right--handed' neutrino) can be accomodated by a set of
$\mathbf 1$, 
$\bar{\mathbf 5}$, and $\mathbf 10$ representations. However, the flipped
and standard versions
of 
$SU(5)$ differ in how the 16 matter fields of each generation are
embedded in these
representations. Flipped $SU(5)$ received its name from the exchanges
in the assignments of the fields:  
up--like and down--like fields are exchanged, as are electron--like 
with neutrino--like, as well as their anti--particle companions.
Thus, in flipped $SU(5)$, the 16 components of a given generation are
distributed as follows among a
set of ${\mathbf 1}$, $\bar{\mathbf 5}$, ${\mathbf 10}$ representations:
${\mathbf 1}_i = \ec{i}$, 
${\bar {\mathbf 5}}_i = \{ \uc{i}, \L{i}\}$,
${\mathbf 10}_i = \{ \Q{i}, \dc{i}, \Nc{i}\}$, where $i=1,2,3$.
This allows
Higgs decuplets to include an electroweak singlet, and the appearance of a
vacuum expectation value (VEV) for this singlet then
breaks $SU(5)$ to $SU(3)_C\times SU(2)_L\times U(1)_{Y}$, rendering
unnecessary adjoint or larger Higgs fields. However, we recall
that the electroweak--doublet Higgs fields $h_u$ and $h_d$ 
of flipped $SU(5)$ appear in standard 
${\mathbf 5}$ and $\bar{\mathbf 5}$ representations.
%flipped $SU(5)$ simply flips the embeddings of $h_u$ and $h_d$ 
%between the two five--plets
%compared to that of their standard $SU(5)$ embeddings.  

String--derived flipped $SU(5)$ was constructed in the 
free--fermion formulation~\cite{fff} of the perturbative heterotic string.
In principle, the superpotential terms in flipped $SU(5)$
or a similar free--fermion model can be calculated to any 
finite order, using the free--fermionic rules for level--one world--sheet
field couplings that were developed some time ago \cite{wsc1,wsc2}. This
has enabled the phenomenology of \fsu to be studied in substantial detail
in this perturbative regime. 
String--derived flipped $SU(5)$ has a characteristic that is
generic to (quasi)--realistic $SU(3)_C\times SU(2)_L\times U(1)_{Y}$ 
or GUT models
with three chiral generations that are of free--fermion, free--boson, or
orbifold construction. Namely, the model contains several supplementary
gauged
Abelian symmetries, one of which, denoted by
$U(1)_A$, is anomalous~\cite{dsw,anoma,cfs}. The
anomaly appears because the trace of the $U(1)_A$ charge operator over the
massless fields is non--zero: $\Tr Q^{(A)}\ne 0$.

The appearance of such an anomalous $U(1)_A$ has profound phenomenological
effects. For instance, in a generic \fsu model, such a $U(1)_A$ imposes
constraints on fermion masses, $R$--violating couplings, and proton decay
operators~\cite{elr}.  Much of the influence of a $U(1)_A$ in string
models results as a by--product of the Green--Schwarz anomaly--cancellation
mechanism and the retention of space--time supersymmetry following the
cancellation.  The latter requires several fields with anomalous charges
to acquire VEVs along a `flat direction', i.e., a direction in field space
with vanishing scalar potential. This alters the classical vacuum of the
model and hence the phenomenology~\cite{anoma,cfs}.  In this paper we
explore field
directions of the \fsu model that are flat to all orders in the
higher--order superpotential terms, and discuss various issues in their
associated phenomenology. In Section 2 we briefly review the meaning of
flat--direction VEVs and their associated $D$-- and
$F$--flatness constraints. Then,
in Section 3 we present the set of all--order flat directions we have
found for string--derived flipped $SU(5)$, along with a discussion how
they were generated.  In Section 4 we consider phenomenological features
of these directions, and compare them with those of other field
directions, whose flatness was proven only up to a finite order. We
conclude our discussion in Section 5. 

\section{Generic Flat Directions}

\subsection{Constraints from $D$-- and $F$--Flatness}

In globally supersymmetric theories, such as the effective field theories
derived from superstring models, there are both $D$ terms,
$D_a^{\alpha}$, and $F$ terms, $F_{\Phi_m}$, contributing to the
scalar potential: 
\beqn
 V(\varphi) = \half \sum_{\alpha} g_{\alpha} 
(\sum_{a=1}^{{\rm {dim}}\, ({\cal G}_{\alpha})} D_a^{\alpha} D_a^{\alpha}) +
                    \sum_m | F_{\Phi_m} |^2\,\, .
\label{vdef}
\eeqn
There is a $D$ term corresponding to each gauge group factor ${\cal
G}_{\alpha}$, and the $D_a^{\alpha}$ in (\ref{vdef}) have the general form
\beqn
D_a^{\alpha}&\equiv& \sum_m \varphi_{m}^{\dagger} T^{\alpha}_a \varphi_m\,\, , 
\label{dtgen} 
\eeqn
where $T^{\alpha}_a$ is a matrix generator of the gauge group ${\cal G}_{\alpha}$ 
for the representation $\varphi_m$.  
For an Abelian gauge group, (\ref{dtgen}) simplifies to
\beqn
D^{i}&\equiv& \sum_m  Q^{(i)}_m | \varphi_m |^2 \label{dtab}\,\,  
\eeqn
where $Q^{(i)}_m$ is the $U(1)_i$ charge of $\varphi_m$. We recall that
$D$ terms originate in the kinetic part of a supersymmetric lagrangian.

We also recall that there is an $F$ term in (\ref{vdef}) for each
superfield $\Phi_{m}$ 
appearing in the superpotential:
\beqn
F_{\Phi_{m}} &\equiv& \frac{\partial W}{\partial \Phi_{m}} \label{ftgen}\,\, . 
\eeqn
Here, the $\varphi_m$ are the scalar--field superpartners     
of the chiral spin--$\half$ fermions $\psi_m$, which together  
form a superfield $\Phi_{m}$.

We recall that, in such a globally supersymmetric theory,
$\vev{V}> 0 $ implies the breaking of space--time supersymmetry. 
Thus, since all of the $D$ and $F$ contributions to (\ref{vdef}) 
are positive semi--definite, each must have 
a zero expectation value in order that $\vev{V}=0$ and
supersymmetry remains unbroken down to a relatively low mass scale. 

An anomalous $U(1)_A$ makes its presence known in
the low--energy effective field theory of a string model 
via triangle diagrams with gauge fields on all three external legs.
Anomalies may appear in these triangle diagrams when
either one or three of the external legs are associated with 
gauge bosons of the anomalous $U(1)_A$. 
In heterotic strings, the entire set of anomalous triangle diagrams is
cancelled by an additional diagram generated by the VEV of the dilaton. 
This also adds to the $D$ term of the anomalous $U(1)_A$ 
a Fayet--Iliopoulos (FI) term:
\beqn
D^{(A)}\equiv \sum_m  Q^{(A)}_m | \varphi_m |^2 + \eps\,;\quad
\epsilon &\equiv& \frac{\gst \MPt}{192 \pi^2} {\rm Tr}\, Q^{(A)}\, ,
\label{epsdef}
\eeqn
where $g_{s}$ is the string coupling and $M_P$ is the reduced Planck mass:
$M_P\equiv M_{Planck}/\sqrt{8 \pi}\approx 2.4\times 10^{18}$ GeV. 
By itself, the FI term would make
a positive--definite contribution to the scalar potential:
$\vev{V} \sim \half g_A \epsilon^2 \,$,
and would break space--time supersymmetry at a scale $\sqrt{\eps}$.
The recovery of supersymmetry requires
a set of scalars to receive VEVs, in such a way that the total scalar VEV
contribution to the
anomalous $D$ term cancels the FI contribution:
\beqn
\vev{D^{(A)}}&\equiv& \sum_m  Q^{(A)}_m | \vev{\varphi_m} |^2 + \eps = 0\, .
\label{dtabc}  
\eeqn
An anomalous $U(1)_A$ therefore induces a shift in the classical vacuum,
while retaining flatness for the 
the non--anomalous Abelian $D$ terms, 
the non--Abelian $D$ terms, 
the $F$ terms, and the superpotential as a whole:
\beqn
\vev{D^{i}}= \vev{D_a^{\alpha}}=0;\quad 
\vev{F_{\Phi_{m}}}= 0;\quad {\rm and}\quad
\vev{W}=0\,.
\label{dft}
\eeqn
The constraints (\ref{dft}) severely limit the set of
scalars that could possibly be chosen non--perturbatively 
so as to satisfy (\ref{dtabc}). 

\subsection{Stringent $F$--Flatness and Non--Abelian Self--Cancellation}

A given $F$ term $F_{\Phi_{m}}$ may contain several components of similar
or various orders $n_i$:
\beqn
\vev{F_{\Phi_m}}&\sim& 
         \sum_i \lambda_{n_i} \vev{\varphi}^2 (\frac{\vev{\varphi}}{\MS})^{n_i-3}\, .
\label{fwnb2}
\eeqn
For a generic $D$--flat set of scalar VEVs, the resulting 
contributions to a given $F$--term will cancel among themselves 
only up to a given order $n_{i'}$. Then $F$--flatness, and thus
supersymmetry, may in general be broken at order $n_{i'+1}$. 
In a particular model, $F$--flatness can often be verified
up to a given order $n_i$ for all $F$ terms, but the exact order at which 
$F$--flatness disappears usually remains undetermined.  
It is clear that, the higher the order to which $F$--flatness is demanded,
the fewer the $D$--flat directions that remain.   

It is also clear that, the lower the order of an $F$--breaking term,
the closer is the scale of supersymmetry breaking to the string scale.
Since the FI scale is about an order of magnitude below the Planck
scale, retention of space--time supersymmetry 
down to the electroweak scale in the
observable sector probably requires $F$--flatness up to about the 17$^{th}$
order in
the weak--coupling limit, and to even higher orders as the coupling
strength increases. For a generic $D$--flat direction, the flatness
of each $F$ term to such a high order
would be extremely difficult to show if component cancellation is
involved. 
However, for a subset of $D$--flat directions this can be 
avoided, and $F$--flatness can be shown to all finite orders.
We term this subset of directions `stringently' $F$--flat.

To be stringently $F$--flat means that each $\vev{F_{\Phi_m}}$ is zero,
not because different components cancel among themselves, but because {\it
each component}
in $\vev{F_{\Phi_m}}$ is individually zero. For an $F$ term containing
only fields with Abelian charges,
stringent flatness holds if
each component of $F_{\Phi_m}$ has one or more fields that 
do not acquire VEVs.
For an $F$ term containing non--Abelian fields, this requirement can be
relaxed slightly.
Because non--Abelian fields contain more than one field component, 
{\it self--cancellation}~\cite{cfn4} of a dangerous $F$ term can
sometimes 
occur along \NA directions. That is, a contraction of two non--Abelian field VEVs
may still be zero. Thus, for some directions it may be possible 
to maintain `stringent' $F$--flatness even when dangerous 
$F$--breaking terms appear in the superpotential derived from string
theory.
 
\section{Flat Directions in Flipped $SU(5)$}

In this Section, we investigate both Abelian (singlet) and non--Abelian
stringently flat directions, along with `self--cancelling' \NA flat 
directions. We start by discussing the
retention of $F$--flatness by self--cancellation in the flipped $SU(5)$
model,
and determine means by which this might be 
implemented,\footnote{Previous flipped $SU(5)$ investigations, such
as~\cite{fsu97} postulated the
self--cancellation of otherwise dangerous $F$ terms.} before moving on to
investigate stringent flatness.

\subsection{Self--Cancellation} 

The full gauge group of the string--derived \fsu model
is 
\beqn
[SU(5)\times U(1)\times \prod_{i=1}^{4} U(1)_i]_{\rm obs} \times
[SO(10)\times SO(6)]_{\rm hid}\, . 
\label{fsu5gg}
\eeqn
Flat directions that cancel
the FI term can be formed from Abelian
fields
carrying only $U(1)_i$ charges or from $SO(10)_{\rm hid}$ and $SO(6)_{\rm
hid}$
fields that are also $SU(3)_C\times SU(2)_L\times U(1)_{Y}\in SU(5)$ singlets. 
Since we shall need many of its aspects, 
for convenience and completeness, the field content of the
string--derived \fsu model is displayed in Tables~\ref{tab:content1}
and~\ref{tab:content2}.
The massless fields of $SO(10)_{\rm hid}$ are five 
fundamental vector $\mathbf 10$'s, denoted by $\T{i=1\,{\rm to}\, 5}$,
while those of $SO(6)_{\rm hid}$ are five
fundamental vector $\mathbf 6$'s, denoted by
$\d{j=1\,{\rm to}\, 5}$, and six pairs of  
$\mathbf 4$ and $\bar{\mathbf 4}$ spinors, denoted by $\a{k=1\,{\rm to}\,
6}$
and $\ba{k=1\,{\rm to}\, 6}$.
Whilst the ${\mathbf 10}$'s and ${\mathbf 6}$'s are $SU(5)\times U(1)$
singlets,
the $\mathbf 4$ and $\bar {\mathbf 4}$'s carry $U(1)$ charge $Q=\pm 5/4$,
resulting in electric charges $Q_{\rm E}=\pm {\half}$. Thus, the
$\mathbf 4$ and $\bar{\mathbf 4}$'s
cannot appear in FI--cancelling flat directions. Rather, it is expected
that 
they form $Q_{\rm E} = Q= 0 $ condensates at an intermediate scale. 
In our treatment of effective bilinear and trilinear terms containing 
${\mathbf 4}\cdot \bar{{\mathbf 4}}$ condensates we assume
the condensation scale to be no higher than 
${\cal O}{(10^{13}\,\, {\rm GeV})}$, and most likely lower, as we 
discuss later.

For the fundamental vector representation of any $SO(2n)$ algebra, 
the $n (2n-1)$ generators of the algebra are imaginary antisymmetric
matrices $M_{a,b}$, with $a,b\in 1~{\rm to}~ 2n$ and $a<b$, of the form:
\beqn
(M_{a,b})_{j,k} = -i (\delta_{a,j}\delta_{b,k} - \delta_{b,j}\delta_{a,k})    
\label{mab}
\eeqn
with commutation relations:
\beqn
[M_{a,b},M_{c,d}] = -i (\delta_{b,c}M_{a,d} - \delta_{a,c}M_{b,d} 
                      + \delta_{a,d}M_{b,c} - \delta_{b,d}M_{a,c} )\, .
\label{cmab}
\eeqn
The Cartan generators form an  $n$--dimensional subset of 
matrices $M_{2c-1,2c}$.
Generic fundamental vector solutions of the entire set of non--linear 
$SO(2n)$ $D$--flat constraints,
\beqn
<D_{a,b}^{SO(2n)}>&\equiv& <\sum_m \varphi_{m}^{\dagger} M_{a,b} \varphi_m> = 0\,\, , 
\label{dtgenb} 
\eeqn
correspond to gauge--invariant products of the vector fields~\cite{gifd}.
For example, we note the following tensor product rules for
low--dimensional representations of $SO(10)$:
\beqn
{\mathbf 10}\times {\mathbf 10} &=& {\mathbf 1}\oplus {\mathbf 45} \oplus {\mathbf 54} \label{tp1}\\      
{\mathbf 10}\times {\mathbf 45} &=& {\mathbf 10}\oplus {\mathbf 120} \oplus {\mathbf 320} \label{tp2}\\      
{\mathbf 10}\times {\mathbf 54} &=& {\mathbf 10}\oplus {\mathbf 210'} \oplus {\mathbf 320} \label{tp3}\\      
{\mathbf 45}\times {\mathbf 45} &=& {\mathbf 1}\oplus {\mathbf 45} \oplus {\mathbf 54}+\dots \label{tp4}\\      
{\mathbf 45}\times {\mathbf 54} &=& {\mathbf 45}\oplus {\mathbf 54} \oplus {\mathbf 210}+\dots \label{tp5}\\      
{\mathbf 54}\times {\mathbf 54} &=& {\mathbf 1}\oplus {\mathbf 45} \oplus {\mathbf 54}+\dots \label{tp6}      
\eeqn 

\newpage

\begin{table}[!ht]
%\begin{flushleft}
\mbox{ \hskip .35in
\begin{tabular}{|l|l|rrrrrrrr|}
\hline
Sector& States &$SU(5)$&$SO(4)$&$SO(10)$&$U(1)$&$U_1$&$U_2$&$U_3$&$U_4$\\    
\hline \hline
\obo  & \oP{1\, {\rm to}\, 5} &    1&     1&      1&     0&    0&    0&    0&    0\\  
      & \op{23}&    1&     1&      1&     0&    0&   -4&    4&    0\\
      &\opb{23}&    1&     1&      1&     0&    0&    4&   -4&    0\\
      & \op{12}&    1&     1&      1&     0&   -4&    4&    0&    0\\
      &\opb{12}&    1&     1&      1&     0&    4&   -4&    0&    0\\
      & \op{31}&    1&     1&      1&     0&    4&    0&   -4&    0\\
      &\opb{31}&    1&     1&      1&     0&   -4&    0&    4&    0\\
      & \oh{1} &    5&     1&      1&    -4&    4&    0&    0&    0\\ 
      &\ohb{1} &   -5&     1&      1&     4&   -4&    0&    0&    0\\ 
      & \oh{2} &    5&     1&      1&    -4&    0&    4&    0&    0\\ 
      &\ohb{2} &   -5&     1&      1&     4&    0&   -4&    0&    0\\ 
      & \oh{3} &    5&     1&      1&    -4&    0&    0&    4&    0\\ 
      &\ohb{3} &   -5&     1&      1&     4&    0&    0&   -4&    0\\ 
\hline
\ob{1}& \oF{1} &   10&     1&      1&     2&   -2&    0&    0&    0\\ 
      &\ofb{1} &   -5&     1&      1&    -6&   -2&    0&    0&    0\\ 
      &\olc{1} &    1&     1&      1&    10&   -2&    0&    0&    0\\ 
\hline
\ob{2}& \oF{2} &   10&     1&      1&     2&    0&   -2&    0&    0\\ 
      &\ofb{2} &   -5&     1&      1&    -6&    0&   -2&    0&    0\\ 
      &\olc{2} &    1&     1&      1&    10&    0&   -2&    0&    0\\ 
\hline
\ob{3}& \oF{3} &   10&     1&      1&     2&    0&    0&    2&   -2\\ 
      &\ofb{3} &   -5&     1&      1&    -6&    0&    0&    2&    2\\ 
      &\olc{3} &    1&     1&      1&    10&    0&    0&    2&    2\\ 
\hline
\ob{4}& \oF{4} &   10&     1&      1&     2&   -2&    0&    0&    0\\ 
      & \of{4} &    5&     1&      1&     6&    2&    0&    0&    0\\ 
      &\olcb{4}&    1&     1&      1&   -10&    2&    0&    0&    0\\ 
\hline
\ob{5}&\oFb{5} &  -10&     1&      1&    -2&    0&    2&    0&    0\\ 
      &\ofb{5} &   -5&     1&      1&    -6&    0&   -2&    0&    0\\ 
      &\olc{5} &    1&     1&      1&    10&    0&   -2&    0&    0\\ 
\hline
\end{tabular}
}
\caption{\it Massless particle states in string-derived flipped
$SU(5)$~\cite{fsu5a}: \obo, \ob{1,2,3,4,5} sectors.}
\label{tab:content1}
%\end{flushleft}
\end{table}

These product rules indicate that several different types of invariants
are possible 
for an even number of $\mathbf 10$'s. 
For two $\mathbf 10$'s, the only invariant in (\ref{tp1}) is a trace product of the
two $\mathbf 10$'s, ${\mathbf 1} = \sum_{i=1}^{10} {\mathbf 10}_i {\mathbf 10}_i$.
However, with four $\mathbf 10$'s, three different invariants can be formed
from the tensor product of two right--hand sides of (\ref{tp1}) since 
${\mathbf 1}\times {\mathbf 1}= 1$,
${\mathbf 45}\times {\mathbf 45}= 1 + \dots$, and 
${\mathbf 54}\times {\mathbf 54}= 1 + \dots$. 
Analogous invariants exist for any $SO(2n)$.

A dangerous $F$ term 
containing VEVs of $SO(10)$ decuplets or $SO(6)$ sextets 
can sometimes be eliminated~\cite{fsu97}
for a given \fsu non--Abelian $D$--flat direction. 
For example, a flat direction could contain four decuplets ${\mathbf 10}^{a =1,4}$
where all VEV

\newpage

\begin{table}[!ht]
%\begin{flushleft}
\mbox{ \hskip .25in
\begin{tabular}{|l|l|rrrrrrrr|}
\hline 
Sector& States &$SU(5)$&$SO(4)$&$SO(10)$&$U(1)$&$U_1$&$U_2$&$U_3$&$U_4$\\    
\hline \hline
$\bS+$
      &\ohb{45}&   -5&     1&      1&     4&    2&    2&    0&    0\\   
$\b{4}+\b{5}$
      & \oh{45}&    5&     1&      1&    -4&   -2&   -2&    0&    0\\ 
      & \op{45}&    1&     1&      1&     0&    2&    2&    4&    0\\ 
      &\opb{45}&    1&     1&      1&     0&   -2&   -2&   -4&    0\\
      &\op{1}  &    1&     1&      1&     0&    2&   -2&    0&    0\\
      &\op{2}  &    1&     1&      1&     0&    2&   -2&    0&    0\\ 
      &\op{3}  &    1&     1&      1&     0&    2&   -2&    0&    0\\
      &\op{4}  &    1&     1&      1&     0&    2&   -2&    0&    0\\
      &\opb{1} &    1&     1&      1&     0&   -2&    2&    0&    0\\
      &\opb{2} &    1&     1&      1&     0&   -2&    2&    0&    0\\
      &\opb{3} &    1&     1&      1&     0&   -2&    2&    0&    0\\
      &\opb{4} &    1&     1&      1&     0&   -2&    2&    0&    0\\
      & \op{+} &    1&     1&      1&     0&    2&   -2&    0&    4\\ 
      &\opb{+} &    1&     1&      1&     0&   -2&    2&    0&   -4\\ 
      & \op{-} &    1&     1&      1&     0&    2&   -2&    0&   -4\\ 
      &\opb{-} &    1&     1&      1&     0&   -2&    2&    0&    4\\ 
$\b{i}$
      & \od{1} &    1&     6&      1&     0&    0&   -2&    2&    0\\ 
$ + 2\alpha + (\bX)$    
      & \od{2} &    1&     6&      1&     0&   -2&    0&    2&    0\\ 
      & \od{3} &    1&     6&      1&     0&   -2&   -2&    0&    2\\ 
      & \od{4} &    1&     6&      1&     0&    0&   -2&    2&    0\\ 
      & \od{5} &    1&     6&      1&     0&    2&    0&   -2&    0\\ 
      & \oT{1} &    1&     1&     10&     0&    0&   -2&    2&    0\\ 
      & \oT{2} &    1&     1&     10&     0&   -2&    0&    2&    0\\ 
      & \oT{3} &    1&     1&     10&     0&   -2&   -2&    0&   -2\\ 
      & \oT{4} &    1&     1&     10&     0&    0&    2&   -2&    0\\ 
      & \oT{5} &    1&     1&     10&     0&   -2&    0&    2&    0\\ 
$\b{1}\pm\alpha$
      & \oa{1} &    1&     4&      1&    -5&   -1&    1&    1&    2\\
      & \oa{2} &    1&     4&      1&    -5&   -1&    1&    1&   -2\\
      & \oa{3} &    1&     4&      1&    -5&   -1&    1&    1&   -2\\
      & \oa{4} &    1&     4&      1&    -5&    1&   -1&    1&   -2\\ 
      & \oa{5} &    1&     4&      1&     5&   -1&   -1&    1&   -2\\
      & \oa{6} &    1&     4&      1&    -5&   -3&    1&   -1&    0\\
      &\oba{1} &    1&    -4&      1&     5&    1&   -1&   -1&   -2\\
      &\obap{2}&    1&    -4&      1&     5&   -1&    1&   -1&   -2\\
      &\obap{3}&    1&    -4&      1&     5&   -1&    1&   -1&   -2\\
      &\oba{4} &    1&    -4&      1&     5&   -1&    1&   -1&    2\\
      &\oba{5} &    1&    -4&      1&    -5&    1&    1&   -1&    2\\
      &\obap{6}&    1&    -4&      1&     5&   -1&    3&    1&    0\\
\hline
\end{tabular}
}
\caption{\it Massless particle states in string-derived flipped
$SU(5)$~\cite{fsu5a}: other sectors.}
\label{tab:content2}
%\end{flushleft}
\end{table}

\noindent
components in ${\mathbf 10}^{a =1,2}$ are $\vev{\alpha}$,  while in 
${\mathbf 10}^{a =3,4}$ five components are $\vev{\alpha}$ and 
another five are
$-\vev{\alpha}$. Self--cancellation would occur in any $F$--term containing exactly one of   
${\mathbf 10}^{a =1,2}$ and one of ${\mathbf 10}^{a =3,4}$.    

\subsection{$D$-- and $F$--Flat Singlet Directions}

The flipped $SU(5)$ model contains
20 fields with non--trivial Abelian charges that are 
singlets of all the non--Abelian gauge group factors, as seen in
Table~\ref{tab:singlets} below.
This set of 20 non--trivial singlets  
can be grouped into ten vector--like pairs, where the two members of
each pair carry exactly opposite charges. 
Four of the 20 fields carry identical sets of $U(1)_i$ charges. 
Thus, for our purposes, the model contains fields with just
seven distinct values of the $U(1)_i$ charges.

\begin{table}[!ht]
%\begin{flushleft}
\mbox{ \hskip 1.8in
\begin{tabular}{|l|rrrr|}
\hline
Vector--Like  &     &      &      &      \\
Singlets        &$U_A$&$U'_1$&$U'_2$&$U'_3$\\    
\hline
\hline 
\oP{12}       &   8 &    0 &  -16 &   -8 \\
\oP{23}       &  12 &    4 &   12 &   12 \\
\oP{31}       & -20 &   -4 &    4 &   -4 \\
\op{45}       &   0 &    4 &   -4 &   24 \\ 
\op{1,2,3,4}  &  -4 &    0 &    8 &    4 \\
\op{+}        &  -8 &    8 &    8 &   -4 \\ 
\op{-}        &   0 &   -8 &    8 &   12 \\ 
\hline
\end{tabular}
}
%\end{flushleft}
\caption{\it The complete set of singlet fields with at least one
non--zero $U(1)_i$ charge, but no non--Abelian charges. The 
normalization of the $U(1)_i$ charges in this paper is four times larger 
than that used in~\cite{fsu5a}.}
\label{tab:singlets}
\end{table}

\vskip .6truecm

The three independent non--anomalous $D$ constraints
result in the four--dimensional (when \op{i} is fixed) 
non--trivial basis set of independent vector--like non--anomalous $D$--flat 
directions shown in Table~\ref{tab:flats}.
To each of these non--trivial directions, elements of a
trivial basis set of $D$--flat directions may be added. This latter set
is composed of the 10 pairs of vector--like fields, 
$(\P{12},\bP{12})$, $(\P{23},\bP{23})$, $(\P{31},\bP{31})$, 
$(\p{45},\bp{45})$, 
$(\p{i},\bp{i})$ for $i\in\{1,2,3,4\}$, $(\p{+},\bp{+})$, 
$(\p{-},\bp{-})$, the three
$(\p{j},\bp{1})$ pairs for $j\in\{2,3,4\}$, and 
the five totally uncharged moduli fields, \oP{1,2,3,4,5}.   

\begin{table}[!ht]
%\begin{flushleft}
\mbox{ \hskip 1in
\begin{tabular}{|l|r|rrrrrrr|}
\hline
Direction        &$Q_A$&\oP{12}&\oP{23}&\oP{31}&\op{45}&\opi  &\op{+}&
\op{-}\\
\hline
\hline 
$\b{\bP{23}} $& -60 &    0  &   -3  &    0  &    1  &   0  &   3  &    2  \\
$\b{\P{31}}$  & -60 &    0  &    0  &    3  &    1  &   0  &   0  &   -1  \\
$\b{\P{12}}$  &   0 &    1  &    0  &    0  &    0  &   0  &   1  &    1  \\
$\b{\p{i}} $  &   0 &    0  &    0  &    0  &    0  &  -2  &   1  &    1  \\
\hline
\end{tabular}
}
\caption{\it Non--trivial basis set of singlet $D$--flat directions.
The numerical entries specify the ratios of the norms of the VEVs of the
fields. A negative entry indicates that the vector 
partner of the field, rather than the field, takes on the VEV. 
Accompanying the $\b{\Phi}$ ($\b{\phi}$) directions are 
the respective vector--partner directions, $\b{\bar{\Phi}}= - \b{\Phi}$
($\b{\bar{\phi}}= - \b{\phi}$).}
\label{tab:flats}
%\end{flushleft}
\end{table}

\vskip .6truecm

We have generated $D$--flat directions $\dvs = \sum_{x} n_x \b{x}$
for integer $n_x$ in the range of $-10$ to $10$ with the 
constraint that $n_{\bp{23}}+n_{\p{31}} > 0$, so that $Q_A \lt 0$.
We tested each of these directions for `stringent' $F$--flatness up 
to at least fifth order in the superpotential. Three directions passed 
this test, with each actually stringently $F$--flat to {\it all finite
orders}, as can be shown simply by gauge invariance constraints.
The three solutions $\dv{1}$, $\dv{2}$, and $\dv{3}$ are given in  
Table~\ref{tab:solutions} below. We note that
$\dv{2}$ is the `root' of the flipped $SU(5)$ flat direction 
analyzed in~\cite{fsu5a},
whilst $\dv{3}$ corresponds to the alternative flat direction studied
in~\cite{fsu97}. Finally,
$\dv{3}$ is a linear combination of the first two, 
$\dv{3} = \dv{1} + 3 \dv{2}$ and actually 
represents an entire two--dimensional class of all--order flat directions,
whose members are linear combinations $\alpha_1 \dv{1} + \alpha_2 \dv{2}$, 
where $\alpha_1$ and $\alpha_2$ are real, positive coefficients. We recall
that
$\dv{1}$, $\dv{2}$, and $\dv{3}$ can be modified by allowing VEVs
for any 
or all of the uncharged moduli fields, $\P{i=1,2,4,5}$,
whilst retaining $F$--flatness to all finite orders. The moduli can take
on VEVs without harming flatness because
the entire set of fields $\{\Pb{23},\P{31},\p{45},\pb{-},\p{+}\}$
is linearly independent with regard to all $U(1)_i$ charges:  
no product of only these fields can ever appear in the superpotential.
Note that $\P{3}$ cannot be appended to $\dv{1,2,3}$, because of the 
renormalizable terms $\P{3} [ \p{+}\pb{+} + \p{-}\pb{-} + \p{45}\pb{45} ]$.
Moreover, it can be shown that a flat direction with simply $\dv{1}$ 
as its root would present some phenomenological problems. Thus, we make
the significant observation that {\it the 
root--space of viable flipped $SU(5)$ singlet flat directions has
been covered in the papers to date.}

\begin{table}[!ht]
%\begin{flushleft}
\mbox{ \hskip 0.2in
\begin{tabular}{|l|r|r|rrrrrrr|}
\hline
Directions & \al \,\,\,\, &$Q_A$ &\oP{12}&\oP{23}&\oP{31}&\op{45}&\opi
&\op{+}& \op{-}\\
\hline
\hline 
$\dv{1}$&$9.2\times 10^{16}$ GeV
              &  -60 &    0  &    0  &    3  &    1  &   0  &   0  &   -1  \\
$\dv{2}$&$9.2\times 10^{16}$ GeV 
              &  -60 &    0  &   -1  &    2  &    1  &   0  &   1  &    0  \\
$\dv{3}$&$4.6\times 10^{16}$ GeV      
              & -240 &    0  &   -3  &    9  &    4  &   0  &   3  &   -1  \\
\hline
\end{tabular}
}
\caption{\it The only $D$--flat directions, mod $\vev{\P{1,2,4,5}}$,
that are $F$--flat to at least fifth order in the superpotential.
These three directions are actually flat to all finite order.}
\label{tab:solutions}
%\end{flushleft}
\end{table}

The non--trivial set of singlet $D$--flat directions can be expanded 
by allowing hidden--sector non--Abelian fields also to acquire VEVs. 
This provides 14 additional basis directions that do not break
the MSSM gauge group. However, we do not include $SO(6)$ 
$\vev{a_i \bar{a}_j}$ condensates among these additional directions,
since their hidden--sector condensation scale should be significantly 
below the FI scale.

\begin{table}[!ht]
%\begin{flushleft}
\mbox{ \tiny
\begin{tabular}{|l|r|rrrrrrrrrrrrrrrrrr|}
\hline
Dir.     &$Q_A$&\opi  &\op{+}& \op{-}&
$\d{1}$&$\d{2}$&$\d{3}$&$\d{4}$&$\d{5}$&
$\T{1}$&$\T{2}$&$\T{3}$&$\T{4}$&$\T{5}$&
$\F{1}$&$\F{2}$&$\F{3}$&$\F{4}$&$\bF{5}$\\
\hline
\hline
$\b{\d{1}}$ & 60&-1&-3&-2&6&0&0&0&0&0&0&0&0&0&0&0&0&0&0\\
$\b{\d{2}}$ & 60&-1& 0& 1&0&6&0&0&0&0&0&0&0&0&0&0&0&0&0\\
$\b{\d{3}}$ & 60& 2&-3& 1&0&0&6&0&0&0&0&0&0&0&0&0&0&0&0\\
$\b{\d{4}}$ & 60&-1&-3&-2&0&0&0&6&0&0&0&0&0&0&0&0&0&0&0\\
$\b{\d{5}}$ &-60& 1& 0&-1&0&0&0&0&6&0&0&0&0&0&0&0&0&0&0\\
\hline
$\b{\T{1}}$ & 60&-1&-3&-2&6&0&0&0&0&6&0&0&0&0&0&0&0&0&0\\
$\b{\T{2}}$ & 60&-1& 0& 1&0&6&0&0&0&0&6&0&0&0&0&0&0&0&0\\
$\b{\T{3}}$ & 30& 1& 0&-1&0&0&6&0&0&0&0&3&0&0&0&0&0&0&0\\
$\b{\T{4}}$ &-60& 1& 3& 2&0&0&0&6&0&0&0&0&6&0&0&0&0&0&0\\
$\b{\T{5}}$ & 60&-1& 0& 1&0&0&0&0&6&0&0&0&0&6&0&0&0&0&0\\
\hline
$\b{\F{1}}$ &  0& 0& 1& 1&0&0&0&0&0&0&0&0&0&0&2&0&0&0&2\\
$\b{\F{2}}$ &  0& 0& 0& 0&0&0&0&0&0&0&0&0&0&0&0&1&0&0&1\\
$\b{\F{3}}$& 0&-1& 1& 0&0&0&0&0&0&0&0&0&0&0&0&0&2&0&2\\
$\b{\F{4}}$ &  0& 0& 1& 1&0&0&0&0&0&0&0&0&0&0&0&0&0&2&2\\
\hline
\end{tabular}
}
\caption{\it Basis set of non--Abelian $D$--flat directions that
leave the MSSM gauge group invariant.
The numerical entries have the same notation as in
Table~\ref{tab:flats}.}
\label{tab:nonAflat}
\end{table}
%\end{flushleft}

\vskip .6truecm

Table~\ref{tab:nonAflat} displays the
basis set of non--Abelian $D$--flat directions that 
leave the MSSM gauge group invariant.
As seen in Table~\ref{tab:nonAflat}, the only components of 
$\F{i}$ and $\bF{5}$ that acquire VEVs in
each of the $\b{\F{i}}$ directions are the respective singlets 
$\nu^c_i$ and $\bar{\nu}_5^c$.
These are the VEVs that break 
$SU(5)\rightarrow SU(3)_C\times SU(2)_L\times U(1)_{Y}$. We note that
$SU(5)$ $D$--flatness requires equal $\nu^c_i$ and $\bar{\nu}_5^c$ VEVs.    
From Table~\ref{tab:nonAflat} we see that Abelian $D$--flatness
independently requires this VEV ratio. This implies that minimally
one $\b{\F{i}}$ basis direction must appear in
a phenomenologically viable flat direction (with which
$SU(5)$ is broken). 

In order to obey the {\it stringent}
$F$--flatness constraints, $\b{\F{2}}$ and $\b{\F{3}}$ are the only possible
choices for an $SU(5)$--charged flat--direction component. 
First note that, in the contraction of two $\bf 10$ (or 
two $\overline{\bf 10}$) representations,
there is an antisymmetrization factor $\epsilon^{ij}$. This implies that
$\vev{\F{i}\cdot\F{i}}=\vev{\Fb{5}\cdot\Fb{5}}= 0$. Hence trilinear terms like
$\F{1}\F{1}\h{1}$ pose no threat to $F$--flatness.
The only relevant trilinear term  is $\bF{5}\cdot \F{4} \p{3}$, which prevents  
an $\F{4}$ VEV.
Analogously, the non--renormalizable fifth--order term
$(\bF{5}\cdot \F{1})^2 \Pb{12}$ prevents an $\F{1}$ VEV.

Terms dangerous for $\F{2}$ and $\F{3}$ VEVs first appear at fifth 
and fourth order, respectively:
\beqn
&&\{ (\bF{5}\cdot \F{2}) \P{31} \T{2}\cdot \T{5},\,\,
     (\bF{5}\cdot \F{2}) \p{2}  \T{2}\cdot \T{4} (1 + \P{1} + \P{5})\}      
\label{bF5F2fb}\\
{\rm and}&&\nolabel\\
&&\{ (\bF{5}\cdot \F{3})^2 \pb{45} \p{+},\,\,
     \bF{5}\cdot \F{3} \d{3}\cdot \d{4} (\bP{23} \p{3} + \P{31}\pb{3})\, ,
\nolabel\\
&&    (\bF{5}\cdot \F{3}) \d{3}\cdot \d{5} 
(1 + \sum_{i=1}^5 \P{i}\bP{i}+ \P{12}\bP{12}+\P{23}\bP{23}+\P{31}\bP{31}+
\nolabel\\
&&\phantom{\bF{5}\cdot \F{3} \d{3}\cdot \d{5} \d{5}}
 \sum_{i=1}^4 \p{i}\pb{i}+ \p{45}\pb{45} + \p{+}\pb{+} + \p{-}\pb{-} )   
\}\, .      
\label{bF5F3fb}
\eeqn
Thus, requiring $<\bF{5}\cdot \F{2}>\ne 0$ along a stringent flat
direction implies 
\begin{itemize}
\item $<\T{2}\cdot\T{5}> = 0$ if $<\P{31}>= 0$ or
      $<\T{2}> = <\T{5}> = 0$ if $<\P{31}>\ne 0$, and
\item $<\T{2}\cdot\T{4}> = 0$ if $<\p{2} >= 0$ or
     $<\T{2}> = <\T{4}> = 0$ if $<\p{2} >\ne 0$.
\end{itemize}
Similarly, 
$<\bF{5}\cdot \F{3}>\ne 0$ implies 
\begin{itemize}
\item  $<\pb{45}> = <\p{+}> = 0$,
\item $<\d{3}> = <\d{5}> = 0$, and
\item $<\d{4}>=0$ 
%or $<\Pb{23} \p{3}>
or $< \P{31}>= 0$ or $<\pb{3}>= 0$. 
\end{itemize}
The $D$--flat basis direction $\b{\F{3}}$ contains $\p{+}$.
This implies that 
some combination of $\b{\P{23}}$, $\b{\Pb{12}}$, $\b{\pb{i}}$,
$\b{\d{1},i}$, $\b{\d{4},i}$, and $\b{\T{1},i}$ 
must be added to $\b{\F{3}}$ 
to eliminate the $\p{+}$ VEV.

{At least 21 $D$--flat non--Abelian directions (and their primed associates) 
remain stringently $F$--flat to all finite order: see
Table~\ref{tab:nAflat} below. One feature key to
the all--order flatness of these directions is the 
specific set of world--sheet charges of the
associated fields. The $\p{45}$, $\p{+}$, $\p{-}$, $\d{3}$, $\T{3}$ (and
their conjugates)
are all Ramond fields carrying $X_{56}$ charge, whilst $\F{2}$ and
$\Fb{5}$ are Ramond 
fields carrying $X_{34}$ charge. The $\P{23}$, $\P{31}$ (and their
conjugates) are Neveu--Schwarz
fields with $X_{12}$ and $X_{34}$ charges, respectively. For many of the
21 directions, several gauge--invariant terms of
relatively 
low order (e.g., sixth through eighth) exist that might
break $F$--flatness. However, only one 
of these terms, namely $\vev{\P{31} \p{45}
\d{3}\cdot\d{3}}\vev{\a{2}\bap{2}}$, satisfies the
picture--changed charge--conservation constraints. All the other terms
contain too many
$X_{56}$ Ramond charges to satisfy the picture--changing constraint.
It was shown in~\cite{wsc1,wsc2}
that the maximum number of identical Ramond $X_{i,i+1}$ charges that can
appear
is $n-2-n_{NS}$, where $n$ is the order of the term and $n_{NS}$ is the
number of Neveu--Schwarz
fields in the term. All but one of the potentially dangerous
gauge--invariant terms contain
more than $n-2-n_{NS}$ $X_{56}$ Ramond fields.

Seven of the $\dv{}$, and the corresponding $\dvp{}$, are flat to
all orders, independent of any constraints.
Following $\vev{\a{i} \ba{j}}$ condensation,   
$F$--flatness of the remaining directions (apart from $\dv{22}$) is
threatened by the sixth--order term
$\vev{\P{31} \p{45} \d{3}\cdot\d{3}}\vev{\a{2}\bap{2}}$.
This term is of no concern if the condensation scale is around 
$10^{10}$ GeV or lower. However, if the condensation scale is above this,
then we must require that
\beqn 
\vev{d_3\cdot d_3}=0\, ,
\label{dd3}
\eeqn
as indicated in the last column of Table~\ref{tab:nAflat}.
However, our rough estimate for the condensation scale 
appears to be in the safe low--scale range,
so that (\ref{dd3}) is unnecessary.

\begin{table}
\begin{flushleft}
\mbox{ \hskip -.25in
\begin{tabular}{|l|r|r|rrrrr|r|r|rrr|l|}
\hline 
Notation   & order  &$Q_A$ &\oP{23}&\oP{31}&\op{45}&\op{+}&
\op{-}&$\d{3}$&$\T{3}$&$F_2$&$F_3$&$\bar{F}_5$&constraints\\
\hline
$\dvpp{1}$&$\infty$&  -60 &    0  &    3  &    1  &    0 &    -1 &    0  &  0    & ( 1 &   0 &         1)& none  \\
$\dvpp{2}$&$\infty$&  -60 &   -3  &    6  &    3  &    3 &     0 &    0  &  0    & ( 1 &   0 &         1)& none  \\
$\dvpp{3}$&$\infty$& -240 &   -3  &    9  &    4  &    3 &    -1 &    0  &  0    & ( 1 &   0 &         1)& none  \\
$\dvpp{4}$&$\infty$& -120 &    0  &    9  &    5  &    0 &    -5 &    0  &  6    & ( 1 &   0 &         1)& none  \\
$\dvpp{6}$&$\infty$&  -60 &   -1  &    3  &    2  &    0 &     0 &    2  &  0    & ( 1 &   0 &         1)& 
                           $\vev{\d{3}\cdot\d{3}}_{\an}=0$ \\   %
$\dvpp{7}$&$\infty$& -120 &   -2  &    5  &    3  &    1 &     0 &    2  &  0    & ( 1 &   0 &         1)& 
                           $\vev{\d{3}\cdot\d{3}}_{\an}=0$ \\
$\dvpp{8}$&$\infty$&  -60 &   -1  &    4  &    3  &   -1 &     0 &    4  &  0    & ( 1 &   0 &         1)& 
                           $\vev{\d{3}\cdot\d{3}}_{\an}=0$ \\ 
$\dvpp{9}$&$\infty$&  -60 &   -3  &    3  &    4  &    0 &     2 &    6  &  0    & ( 1 &   0 &         1)&
                           $\vev{\d{3}\cdot\d{3}}_{\an}=0$ \\
$\dvpp{10}$&$\infty$&-240 &   -3  &   12  &    7  &    0 &    -1 &    6  &  0    & ( 1 &   0 &         1)& 
                           $\vev{\d{3}\cdot\d{3}}_{\an}=0$ \\
$\dvpp{11}$&$\infty$& -60 &   -3  &    2  &    3  &    3 &     0 &    0  &  4    & ( 1 &   0 &         1)& none  \\
$\dvpp{12}$&$\infty$& -60 &   -3  &    5  &    6  &    0 &     0 &    6  &  4    & ( 1 &   0 &         1)&  
                           $\vev{\d{3}\cdot\d{3}}_{\an}=0$ \\             
$\dvpp{13}$&$\infty$&-420 &   -3  &   18  &   10  &    3 &    -1 &    6  &  0    & ( 1 &   0 &         1)& 
                           $\vev{\d{3}\cdot\d{3}}_{\an}=0$ \\ 
$\dvpp{14}$&$\infty$& -60 &   -3  &    6  &    7  &   -3 &     2 &   12  &  0    & ( 1 &   0 &         1)&
                           $\vev{\d{3}\cdot\d{3}}_{\an}=0$ \\
$\dvpp{15}$&$\infty$& -60 &   -3  &    3  &    4  &    3 &    -1 &    0  &  6    & ( 1 &   0 &         1)& none  \\  
%( 4,    1)
$\dvpp{16}$&$\infty$& -60 &   -3  &    6  &    7  &    0 &    -1 &    6  &  6    & ( 1 &   0 &         1)& 
                           $\vev{\d{3}\cdot\d{3}}_{\an}=0$ \\
%( 4,    3)     
$\dvpp{17}$&$\infty$& -60 &   -2  &    3  &    3  &    1 &     0 &    2  &  2    & ( 1 &   0 &         1)&
                           $\vev{\d{3}\cdot\d{3}}_{\an}=0$ \\
%( 4,    6)
$\dvpp{18}$&$\infty$& -60 &   -2  &    5  &    5  &   -1 &     0 &    6  &  2    & ( 1 &   0 &         1)&
                           $\vev{\d{3}\cdot\d{3}}_{\an}=0$ \\
%( 4,    4)
$\dvpp{19}$&$\infty$&-120 &   -6  &    9  &   11  &    0 &     1 &   12  &  6    & ( 1 &   0 &         1)&
                           $\vev{\d{3}\cdot\d{3}}_{\an}=0$ \\
%( 4,    2)
$\dvpp{20}$&$\infty$& -60 &   -6  &    6  &   10  &    3 &    -1 &    6  & 12    & ( 1 &   0 &         1)&
                           $\vev{\d{3}\cdot\d{3}}_{\an}=0$ \\
%( 4,    5)
$\dvpp{21}$&$\infty$& -60 &   -6  &    9  &   13  &   -3 &     2 &   18  &  6    & ( 1 &   0 &         1)&
                           $\vev{\d{3}\cdot\d{3}}_{\an}=0$ \\
\hline
$\dv{22}$&  8     &  -60 &    0  &    3  &    1  &    0 &    -1 &    0  &  0    &   0 &   2 &         2& none   \\
\hline
\end{tabular}
}
\caption{\it Non--Abelian $D$--flat directions, 
mod $\langle\Phi_{1,2,4,5}\rangle$,
that are $F$--flat to all orders
(with the exception of $\dv{22}$) 
in the superpotential. The scale of the VEV for each of these directions 
is $\vev{\alpha}/\sqrt{-Q_A/60}$, where 
$\vev{\alpha}\equiv 9.2\times 10^{16}$ GeV. 
The primed and unprimed versions of a flat direction vary only by the
presence or absence, respectively, of a $\F{2}\cdot\bF{5}$ VEV component.
In the constraints column, the subscript ``$\an$'' denotes that
$\vev{\d{3}\cdot\d{3}}=0$ is required only if the hidden--sector $SO(6)$
condensation scale of quadruplet $a$ fields 
occurs at or above $\sim 10^{10}$ GeV.}
\label{tab:nAflat}
\end{flushleft}
\end{table}

With the exception of $\dv{22}$, for every direction not containing
$\F{2}$ and $\bF{5}$ VEVs in Table~\ref{tab:nAflat}, there is another that
does. The corresponding directions are
denoted $\dv{i}$ and $\dvp{i}$, respectively, and are contained in the
same row in Table~\ref{tab:nAflat}. Each of these flat directions may 
additionally contain any or all of the uncharged moduli fields $\P{1,2,4,5}$.
The more realistic of our flat directions are clearly 
those in the $\dvpn$ class, since the
breaking of $SU(5)\times U(1)$ to the Standard Model
requires at least one $\vev{\F{i}}\ne 0$ or $\vev{\bF{5}}\ne 0$,
and $SU(5)$ $D$--flatness then requires 
$\vev{\Fb{5}} = \vev{F}$, where $F\equiv \sum_{i=1}^{4} \alpha_i \F{i}$,
for $| \vec{\alpha} | = 1$. 
Whilst an $\F{2}$ VEV was considered in~\cite{rt90}, most recent papers
have considered a VEV for $\F{1}$, rather than $\F{2}$.
Thus, these $\F{2}$ directions possess somewhat different 
phenomenology from those generally investigated, and we therefore 
consider them in the next section, and 
compare our results to those of~\cite{rt90}.

We note that there appears to be no all--order
flat direction containing $\F{3}$: for example, the
flatness of $\dv{22}$ is broken at eighth order by the
superpotential term 
$\vev{\P{31} \bp{45} \bp{-} (\F{3}\cdot\Fb{5})^2} \P{23}$. 
In any case, the flipped $SU(5)$ doublet--triplet splitting
mechanism prevents $\F{3}$ from being alone among the $\F{i}$
fields to acquire a VEV~\cite{fsu5a,rt90}.

\section{Flat--Direction Phenomenology}

From Table~\ref{tab:nAflat} we observe that each of the
all--order flat directions ${\bf d}^{(')}_{1}$ through
${\bf d}^{(')}_{19}$ can be 
embedded in either $\dvp{20}$ or $\dvp{21}$. 
Thus, in the next
subsection we examine the Higgs mass eigenstates and eigenvalues 
resulting from $\dvp{20}$ or $\dvp{21}$. The corresponding 
eigenstates and eigenvalues
for the 19 embedded directions can be easily determined from these
results. 
In the following subsection, we then examine 
for $\dvp{20}$ and $\dvp{21}$ the corresponding masses of 
the three Standard Model generations of quarks and leptons. 
%which are related in any given flat direction.

When we indicate the components of a mass matrix, we generally list only
the leading term, or one representative of them if there are several
leading--order terms. For terms involving $SO(6)$ condensates, $\vev{a_i
\bar{a}_j}$, we assume a condensation scale no higher than $10^{13}$ GeV.
Relatedly, we assume a suppression factor of
$\sim\frac{1}{10^8}$ {\it or less}, rather than $\sim\frac{1}{100}$, for
each condensate. We include  up to eleventh (seventh) order
terms in the mass matrices when condensates are absent (present). 

\subsection{Higgs Mass Textures}

{We first determine the Higgs mass eigenstates and eigenvalues
produced by our all--order flat directions.
As in~\cite{fsu99a}, our $5 \times 5$ Higgs mass matrices   
contain terms for both the $SU(2)_L$ doublet 
and $SU(3)_C$ triplet components of  
the $SU(5)$ $5$ and $\bar{5}$ Higgs representations.  The
$4 \times 4$ $SU(2)_L $ doublet Higgs matrix 
excludes the $\bF{5}$ and $\F{2}$ components, whilst
the $SU(3)_C$ triplet matrix is the entire $5 \times 5$ matrix.
In the absence of $SO(6)$ condensates, it takes the following form: 
%{\smallfont
\begin{equation}
M_{11}  =  \bordermatrix{ 
       &\bh{1}  &\bh{2}&\bh{3}&\bh{45}&\F{2}    \cr
\h{1}  & 
       &   
       &   
       &  
       & <X^{15}_{(0)}>+ \cr
       &  
       &  
       &  
       &  
       & <X^{15}_{(1)}>\pps\cr
       &&&&&\cr
\h{2}  &
       &   
       &   
       &
       & <\F{2}>\cr
       &
       &  
       &  
       &
       & \cr
       &&&&&\cr
\h{3}  & <\P{31}>  
       & <\bP{23}> 
       & 
       &  
       & \cr 
       &   
       &  
       &  
       &  
       & \cr 
       &&&&&\cr
\h{45} &  
       &  
       & <\p{45}>
       &  
       & <X^{45}>
       & \cr
       &&&&&\cr
       &&&&&\cr
\bF{5} & <X^{51}>+       
       & <\bF{5}>  
       &
       & <X^{54}_{(0)}>+ 
       &       \cr 
       & <X^{51}_{(0)}>\pps       
       &   
       &
       & <X^{54}_{(1)}>\pps 
       &       \cr}, 
\label{higgst}
\end{equation}
where
\beqn
X^{15}_{(0)} &=& \F{2} \bP{23} \p{45} \T{3}\cdot \T{3} 
                       (\bp{-}  +  \bP{23} \p{45} \d{3}\cdot\d{3} )    \nolabel\\ 
X^{15}_{(1)} &=& \F{2} (\bP{23} \p{45}\bp{+} \d{3}\cdot \d{3} 
                   +\bP{23}^2\p{45}^2 \d{3}\cdot\d{3}\T{3}\cdot\T{3})\nolabel\\
X^{45}       &=& \F{2}\bP{23} \p{45}           \nolabel\\
X^{51}       &=& \bF{5} \P{31}^2\p{45}^2\d{3}\cdot\d{3}\T{3}\cdot\T{3}
                 \nolabel\\
X^{51}_{(0)} &=& \bF{5} \P{31}\p{45}\p{+}\d{3}\cdot\d{3}\T{3}\cdot\T{3}
                 \nolabel\\
X^{54}_{(0)} &=& \bF{5} \p{+}\T{3}\cdot{\T{3}}   \nolabel\\
X^{54}_{(1)} &=& \bF{5} \p{-}\d{3}\cdot{\d{3}}\, .   \label{xset}
\eeqn

The matrix (\ref{higgst}) contains all condensate--free Higgs mass terms 
for flat directions $\dvp{20}$ and $\dvp{21}$.
If no subscript appears on a term
in (\ref{higgst}), then that term is produced by both flat directions,
On the other hand, 
a subscript $(0)$ or $(1)$ indicates a term that is produced only
by $\dvp{20}$ or $\dvp{21}$, respectively.
Although not explicitly shown,   
the appropriate $n^{\rm th}$--order coupling constants $\lambda_n$ and related
$1/\MS^{n-3}$ factors for each term are implied.

We now exhibit numerical order--of--magnitude estimates of 
the entries in (\ref{higgst}):
\beqn
M^{oom}_{11} &=&  
\bordermatrix{
       &\bh{1}  &\bh{2}&\bh{3}&\bh{45}&\F{2}    \cr
\h{1}  & 0      & 0      &  0     & 0        &10^{-4} \cr
\h{2}  & 0      & 0      &  0     & 0        & 1      \cr
\h{3}  & 1      & 2      &  0     & 0        & 0      \cr
\h{45} & 0      & 0      &  2     & 0        & 10^{-1}\cr
\bF{5} &10^{-4} & 1      &  0     & 10^{-2}  &        \cr}, 
\label{esthiggst}
\eeqn
whose consequences we explore later.

In the presence of $SO(6)$ condensates, there are additional Higgs mass
terms, as follows:
%{\smallfont
\begin{equation}
M^{a\bar{a}}_{7}  =  \bordermatrix{ 
       &\bh{1}  &\bh{2}&\bh{3}&\bh{45}&\F{2}    \cr
\h{1}  & <Y^{11}_{(0)}> 
       &   
       &   
       & <Y^{14}> 
       & <Y^{15}> \cr
       &&&&&\cr
\h{2}  &
       & <Y^{22}>  
       &   
       &
       & \cr 
       &&&&&\cr
\h{3}  &   
       &  
       & <Y^{33}_{(0)}> 
       &  
       & <Y^{35}> \cr 
       & \cr 
       &&&&&\cr
\h{45} & <Y^{41}> + 
       & <Y^{42}> + 
       & 
       & <Y^{44}> + 
       & \cr
       & <Y^{41}_{(0)}>\pps
       & <Y^{42}_{(0)}>\pps
       & 
       & <Y^{44}_{(0)}>\pps
       & \cr
       &&&&&\cr
\bF{5} &        
       &   
       &
       &  
       &       \cr}, 
\label{yhiggst}
\end{equation}
}
where
\beqn
Y^{11}_{(0)} &=& \bF{5}\F{2}\p{+}\a{2}\bap{3} \nolabel\\ 
Y^{14}       &=& \d{3}\cdot \d{3} \a{2} \bap{2}  \nolabel\\          
Y^{15}       &=& \F{2} \a{6} \bap{6}  \nolabel\\
Y^{22}       &=& \P{31}\a{1}\a{1}\a{5}\a{5}   \nolabel\\
Y^{33}_{(0)} &=& \bF{5}\F{2}\p{+}\a{2}\bap{3}   \nolabel\\       
Y^{35}       &=& \F{2} \bap{2} \bap{2} \ca{5} \ca{5}   \nolabel\\
Y^{41}       &=& \p{45}\ca{1}\ca{1}\ca{5}\ca{5}   \nolabel\\
Y^{41}_{(0)} &=& \P{31}\p{45}\p{+}\a{2}\bap{2}    \nolabel\\ 
Y^{42}       &=& \p{45}\bap{2}\bap{2}\ca{5}\ca{5}   \nolabel\\
Y^{42}_{(0)} &=& \bP{23}\p{45}\p{+}\a{2}\bap{2}    \nolabel\\
Y^{44}       &=& \P{31}\a{1}\a{1}\a{5}\a{5}   \nolabel\\  
Y^{44}_{(0)} &=& \p{+}\a{2}\bap{2}\, .         \nolabel     
\eeqn
and the following is
a numerical estimate of (\ref{higgst}) and (\ref{yhiggst}) combined:
\beqn
M^{oom'}_{11} &=& M^{oom}_{11} + M^{a\bar{a},oom}_{7}\nolabel\\    
             &=&  \bordermatrix{
       &\bh{1}  &\bh{2}&\bh{3}&\bh{45}&\F{2}    \cr
\h{1}  &\lsim 10^{-9}_{(0)}  
                &   0        
                         &  0        
                                  & \lsim 10^{-8}           
                                              &10^{-4}\cr
\h{2}  & 0      & 10^{-15}   
                         &  0      
                                  &    0      & 1 \cr
\h{3}  & 1      &2       &\lsim 10^{-9}_{(0)}          
                                  &    0
                                              & \lsim 10^{-15}_{(1)}       \cr
\h{45} &\lsim 10^{-9}_{(0)}  + 10^{-15}_{(1)}   
                &\lsim 10^{-9}_{(0)}+ 10^{-15}_{(1)}  
                         & 2
                                  & \lsim 10^{-7}_{(0)}+ 10^{-15}_{(1)}              
                                              & 10^{-1}      \cr
\bF{5} &10^{-5} &1       & 0      & 10^{-2}   &              \cr}.\nolabel\\ 
\label{yesthiggst}
\eeqn
where we recall that a subscript $(0)$ or $(1)$ indicates a term that is
produced only
by $\dvp{20}$ or $\dvp{21}$, respectively.  

The Higgs doublet matrix embedded in (\ref{esthiggst})  
has the generic form~\footnote{This mass texture 
is embedded in all of the flipped $SU(5)$ Higgs matrices 
in~\cite{fsu99a}.}:
\beqn
M^{gen}_{11}  &=&  \bordermatrix{
       &\bh{1}  &\bh{2}&\bh{3}&\bh{45}       \cr
\h{1}  & 0      & 0      &  0     & 0        \cr
\h{2}  & 0      & 0      &  0     & 0        \cr
\h{3}  & c      & d      &  0     & 0        \cr
\h{45} & 0      & 0      &  f     & 0        \cr}\, , 
\label{genhigg}
\eeqn
for all flat directions in Table~\ref{tab:nAflat}.
We see that (\ref{genhigg}) produces two pairs of 
FI--scale massive doublet eigenstates,
and two pairs of massless Higgs doublet eigenstates.
The (unnormalized) massive Higgs eigenstates are
%\beqn
$h^M_1       \equiv \h{3}$
%\quad {\rm and}\quad
and
$\bar{h}^M_1 \equiv c\, \bh{1} + d\, \bh{2}$
%\label{hbhes1}
%\eeqn
with $M^2= c^2 + d^2$, and
%\beqn
$h^M_2       \equiv \h{45}$
%\quad {\rm and}\quad
and
$\bar{h}^M_2 \equiv \bh{3}$
%\,  \label{bhes2}
%\eeqn
with $M^2= f^2$. 
The massless Higgs are
%\beqn
$\h{1}, \h{2}, d\, \bh{1} - c\, \bh{2}$, and 
$\bh{45}$.
%\, . \label{bh1}
%\eeqn

In (\ref{yesthiggst}) we
note that along the $\bF{5}$ row, and 
$\bh{1}$, $\bh{2}$, and $\bh{45}$ columns, and along the
$\F{2}$ column, and $\h{1}$, $\h{2}$, and $\h{45}$ rows, the triplet 
mass components are non--zero. 
A generic matrix of the form
\beqn
M^{gen,3}_{11}  &=&  \bordermatrix{
       &\bh{1}  &\bh{2}&\bh{3}&\bh{45}&\F{2} \cr
\h{1}  & 0      & 0      &  0     & 0 & k    \cr
\h{2}  & 0      & 0      &  0     & 0 & l    \cr
\h{3}  & c      & d      &  0     & 0 & 0    \cr
\h{45} & 0      & 0      &  f     & 0 & m    \cr 
\bF{5} & g      & i      &  0     & j & 0    \cr}. 
\label{genhiggb}
\eeqn
produces exactly one massless triplet/anti--triplet pair:
%\beqn
$h^{[3], 0} \equiv l\, \h{1} - k\, \h{2}$ 
%\quad {\rm and}\quad
and
$\bar{h}^{[3], 0} \equiv d\, \hb{1} - c\, \hb{2} + \frac{-dg+ci}{j}\, \hb{45}$.
%\label{tbh1}
%\eeqn
Thus, additional terms beyond those in (\ref{higgst})
must appear in the Higgs doublet matrix, 
to provide FI--scale masses for one additional
pair of doublets and the remaining triplet/anti--triplet pair.
As (\ref{yesthiggst}) and (\ref{yhiggst}) indicate,
our flat directions do indeed yield additional terms
when terms containing $SO(6)$ condensates are included. However, the 
contributions from these terms are substantially smaller, since the
condensate
scale would most naturally be no higher than about $10^{13}$ GeV. 
That is, each condensate $\vev{a_i \bar{a}_j}$ contributes a
suppression factor of order $10^{-8}$ or smaller. We have listed in 
(\ref{yhiggst}) the related condensate terms through seventh order.
We find these mass contributions associated with the flat direction
$\dvp{20}$ 
have values $\lsim 10^{-9}$ times the FI scale, namely $\sim 10^{8}$ GeV, 
except for a 
$\h{1}\hb{45}$ component, denoted $Y^{14}$, which is larger 
by an estimated factor $\sim 10$  
and a $\h{45}\hb{45}$ component, denoted $Y^{44}_{(0)}$, 
which is estimated to be larger than these by a
factor $\sim 100$. 
Whilst $\dvp{21}$ also produces the term $Y^{14}$, the remaining
$\dvp{21}$ condensate terms only have   
values $\sim 10^{-15}$ times the FI scale, namely $\sim 100$ GeV.
Thus, the condensate terms unique to the $\dvp{21}$--class directions
can clearly be ignored.

We can effectively ignore all condensate terms in (\ref{yesthiggst}),
except for $Y^{14}$ and $Y^{44}_{(0)}$.
The result of adding both of these terms to $M_{11}$ for $\dvp{20}$,
or only the first for $\dvp{21}$, is to produce a matrix 
of similar form to matrices (20) and (25) of
\cite{fsu99a}. The massless eigenstates for a matrix of the form
\beqn
M^{gen'}_{11}  &=&  \bordermatrix{
       &\bh{1}  &\bh{2}&\bh{3}&\bh{45}       \cr
\h{1}  & 0      & 0      &  0     & p        \cr
\h{2}  & 0      & 0      &  0     & 0        \cr
\h{3}  & c      & d      &  q     & 0        \cr
\h{45} & 0      & 0      &  f     & 0        \cr},
\nolabel\\ 
\label{genhiggc}
\eeqn
are (assuming real VEVs) clearly:
%\beqn
$h\equiv \h{2}$
%  \quad {\rm and}\quad
and
$\bh{}\equiv d\, \bh{1} - c\, \bh{2}$,
Here $d=\vev{\Pb{23}}$ and $c=\vev{\P{31}}$. 
% \, .   
%\label{obh1a}
%\eeqn 

We note that $Y^{14}\sim 10^{-8}$ of the FI scale gives a formerly
massless Higgs doublet pair an intermediate mass
$\sim 10^{8\,\,{\rm or}\,\, 9}$ GeV.
Thus, depending on the condensate scale, it appears that a single 
massless Higgs doublet pair can be produced by some all--order 
flat directions. However, again comparing 
(\ref{genhiggc}) with the matrices of \cite{fsu99a}, we see that
the absence of
a non--zero $\h{1}\bh{1}$ or $\h{1}\bh{2}$ term in (\ref{genhiggc}) 
eliminates an $\hb{45}$
component in $\hb{}$. This has profound phenomenological
consequences that we shall discuss in the following subsection. 

%On the other hand $\dvp{21}$  requires $\vev{\T{3}\cdot\T{3}}=0$
%for all--order flatness and allows $\vev{\T{3}\cdot\T{3}}\ne 0$.
%This is consistent with both the set of $X_{(0)}$ terms and non--zero
%$Y^{14}$. Note also that with $\dvp{21}$, the rational for ignoring 
%all other $Y$--terms, except $Y^{15}$, is without question, while
%for $\dvp{20}$, some other $Y$--terms may only be less than
%$Y^{14}$ by a factor of around 10.      

Recall that in Table~\ref{tab:nAflat} we indicated that for all--order 
flatness, directions $\dvp{20}$ and $\dvp{21}$ may require  
$\vev{\d{3}\cdot\d{3}}=0$, to eliminate 
the appearance of a possibly dangerous $\vev{W}$--term, 
$\vev{\P{31} \p{45} \d{3}\cdot\d{3}}\vev{\a{2}\bap{2}}$. 
Since the $Y^{14}$ term contains $\d{3}\cdot\d{3}$, 
this zero--VEV constraint has severe phenomenologically consequences.
If $\vev{\d{3}\cdot\d{3}}=0$ then $Y^{14}=0$ and, therefore, 
two Higgs doublets and one Higgs triplet would remain massless.
Furthermore, note also that the potentially dangerous $\vev{W}$ term
and $Y^{14}$ 
have the same last four VEV components: 
$\vev{\d{3}\cdot\d{3}}\vev{\a{2}\bap{2}}$. 
Thus, if $\vev{\a{2}\bap{2}}$ has a magnitude low enough that the $W$ term 
can be ignored, then one would expect that the Higgs mass term $Y^{14}$ can 
likewise be ignored.
Alternatively, if $\vev{\a{2}\bap{2}}$ is too large then 
we must require $\vev{\d{3}\cdot\d{3}}= 0$ and the
$\vev{W}$--term and $Y^{14}$ should both vanish.

We observe that, along the 21 
flat directions, only one $SO(6)$ $\bmit 4$--$\bar {\bmit 4}$ 
pair gains a mass at the FI scale. 
For a generic $SU(N_c)$ gauge group containing $N_f< 2 N_c$ flavors 
of massless matter states in vector--like pairings,
$T_i \overline{T}_i$, $i= 1,\, \dots\, N_f$,
the gauge coupling $g_s$
becomes strong at a condensation scale defined by 
$\Lambda = \MP {\rm e}^{8 \pi^2/\beta g_s^2}$, where
the $\beta$--function is given by
$\beta = - 3 N_c + N_f$.
Thus, for $N_c= 4$ and $N_f= 5$,   
$\beta = -7$ and the $SO(6)_H$ condensate scale should be around
$\Lambda = {\rm e}^{-22.5}\MP \sim 4\times 10^8$  GeV.

In the following subsection, however, we briefly explore
some of the 
phenomenology resulting from the Higgs pair $\h{2}$ and 
$\vev{\Pb{23}}\, \hb{1} - \vev{\P{31}}\, \hb{2}$,
under the assumption that they are the only masslesss  
Higgs doublets, 
concentrating on the textures of the quark and charged--lepton mass matrices. 

\subsection{Quark and Charged--Lepton Mass Textures}

In combination with the flat direction $\dvp{20}$ or $\dvp{21}$, 
the massless pair of Higgs fields $\h{2}$ and 
$\vev{\Pb{23}}\bh{1} - \vev{\P{31}}\, \bh{2}$
produce several MSSM quark and lepton mass terms.
However, most of these terms contain $SO(6)$ condensates,
which would most likely result in over--suppression of the 
lower--generation masses (except perhaps for Dirac neutrino 
terms). Hence, for quark and lepton masses,  
we consider only mass terms for which condensates are absent.
Through eighth order these terms are~\footnote{Recall that in our 
all--order flat directions $\F{2}$ is massive at the FI scale. Hence all
terms containing $\F{2}$ decouple.}:
\beqn
{\rm up:} & &     {\rm no~terms},
\label{upterms}\\
{\rm down:} & &   (\F{1} \F{1} + \F{4} \F{4})
\h{2}\vev{\P{31} \p{45} (\p{+} \T{3} \T{3} + \p{-} \d{3} \d{3})}\, ,
\label{doterms}\\
{\rm electron:} & &   
(\fb{2} \lc{2} + \fb{5} \lc{5})  \h{2} +
(\fb{2} \lc{5} + \fb{5} \lc{2}) \h{2} \vev{\Fb{5}\cdot \F{2}}\, .
\label{elterms}
\eeqn
Clearly this is not a viable set: 
No up--quark mass terms appear below at least ninth order.  
Unsuppressed or slightly suppressed up--quark mass terms
only appear when $\bar{h}$ contains a $\hb{45}$ component.  
Specifically, $\hb{45}$ produces viable top and charm masses
from third-- and fifth--order terms, respectively: 
$\F{4} \fb{5} \bh{45}$ and  $\F{4} \fb{2} \bh{45} \vev{\Fb{5}\cdot \F{2}}$.
Along flat direction $\dvp{20}$, an $\hb{45}$ component could be possible 
only if condensates in some specific terms in (\ref{yesthiggst}) 
receive sufficiently large VEVs, e.g., of order $10^{-9}$ or greater. 
Note also that
the down--quark mass matrix has two equivalent fifth--order mass terms, 
for $\F{1} \F{1}$ and $\F{4} \F{4}$. This produces the
further phenomenological disaster of 
equal bottom and strange masses. 
A generic degeneracy of second-- and third--generation down--quark masses 
for $\vev{\F{1}}= 0$, $\vev{\F{2}}\ne 0$
was first noted in \cite{rt90}.  

\section{Concluding Discussion}

Our main result has been to demonstrate
that in the flipped $SU(5)$ model Higgs mass textures 
produced by all--order stringently--flat directions, i.e.,
those where cancellations between components of a given $F$ term 
are not postulated,
are extremely constrained. Generally, two out of four pairs of MSSM Higgs
doublets receive FI--scale masses and so
decouple from the low--energy effective field theory. 
However, along some of our all--order flat directions it
may be possible for three out of the four pairs of Higgs doublets
$h_i$ and $\hb{i}$ to gain FI--scale masses, while one combination remains
massless. Whether or not one or two pairs of Higgs doublets
remain massless appears to depend on the hidden sector $SO(6)$ 
condensate scale. We have also found that,
along our all--order flat directions, the surviving 
$\bar{h}$ will not contain an $\hb{45}$ component,
unless some terms containing 
$\vev{\a{i} \ba{j}}$ condensates appear in the mass matrix. 
However, we recall that the presence in $\bar{h}$ of
an $\hb{45}$ component is critical for a viable top-quark mass term. 

The form of the quark and lepton mass matrices is heavily restricted
for all--order stringent flat directions, and not very realistic.
{\it This reinforces the phenomenological necessity of studying
non--stringently flat directions, wherein supersymmetry is
almost inevitably broken at some finite order.} 
This might even be a positive
advantage, if the breaking occurs at a sufficiently high order.
Thus, building on the analysis started here, 
in~\cite{cen2} we will review the non--stringently--flat directions
investigated previously in \cite{fsu90,fsu97,fsu98,fsu99a},
and determine the respective orders at which $F$--flatness 
is broken for these directions, as well as address other phenomenological 
issues.

\section{Acknowledgments}
The work of G.C. and D.V.N. is supported in part
by DOE Grant No.\  DE--FG--0395ER40917.
\newpage

\bigskip
\medskip

\def\bibiteml#1#2{ }
\bibliographystyle{unsrt}

\hfill\vfill\eject
\end{document}